\documentclass[journal=nalefd,manuscript=letter]{achemso}
\usepackage{graphicx}
\usepackage{float}
\usepackage{color}
\usepackage{bigints}
\usepackage[normalem]{ulem}
\usepackage{bm}
\usepackage{amsmath,amsfonts,amssymb} % mathtools
\usepackage{latexsym}
\usepackage{lmodern}
\usepackage[T1]{fontenc}
\usepackage[english,activeacute]{babel}
\usepackage{mathtools}
\usepackage{color} % color package
\usepackage{xcolor} % color package

\definecolor{ACcol1}{RGB}{0,112,112}
\definecolor{ACcol2}{RGB}{153,0,153}

\usepackage{amssymb} %letras chulis
\usepackage{dsfont} % para escribir la matrix identidad chula

\usepackage{pbsi}

\usepackage[colorlinks,linkcolor=blue,citecolor=blue,urlcolor=blue]{hyperref}

\addto\extrasenglish{}
\addto\extrasenglish{}

\newcommand{\ii}{\mbox{i}}

\newcommand{\LL}{\mbox{\scriptsize{L}}}

\newcommand{\EE}{\mbox{\scriptsize{E}}}

\newcommand{\g}{\mbox{g}}

\newcommand{\rr}{\mbox{\scriptsize{rad}}} % radiative
\newcommand{\nr}{\mbox{\scriptsize{nrad}}} % non-radiative
\setkeys{acs}{keywords = true}

\title{Plexcitonic quantum light emission from nanoparticle-on-mirror cavities}

\author{R. S\'aez-Bl\'azquez}
\email{rocio.saez.blazquez@tuwien.ac.at}
\affiliation{Departamento
de F\'isica Te\'orica de la Materia Condensada and Condensed
Matter Physics Center (IFIMAC), Universidad Aut\'onoma de
    Madrid, 28049 Madrid, Spain}
\alsoaffiliation{Vienna Center for Quantum Science and Technology,
Atominstitut, TU Wien, 1040 Vienna, Austria}

\author{A. Cuartero-Gonz\'alez}
\affiliation{Departamento de F\'isica Te\'orica de la Materia
Condensada and Condensed Matter Physics Center (IFIMAC),
Universidad Aut\'onoma de
    Madrid, 28049 Madrid, Spain}
\alsoaffiliation{Mechanical Engineering Department, ICAI,
Universidad Pontificia Comillas, 28015 Madrid, Spain}

\author{J. Feist}
\affiliation{Departamento de F\'isica Te\'orica de la Materia
Condensada and Condensed Matter Physics Center (IFIMAC),
Universidad Aut\'onoma de
    Madrid, 28049 Madrid, Spain}

\author{F. J. Garc\'ia-Vidal}
\affiliation{Departamento de F\'isica Te\'orica de la Materia
Condensada and Condensed Matter Physics Center (IFIMAC),
Universidad Aut\'onoma de
    Madrid, 28049 Madrid, Spain}
\alsoaffiliation{Institute of High Performance Computing, Agency
for Science, Technology, and Research (A*STAR), Singapore 138632,
Singapore}

\author{A. I. Fern\'andez-Dom\'inguez}
\email{a.fernandez-dominguez@uam.es} \affiliation{Departamento de
F\'isica Te\'orica de la Materia Condensada and Condensed Matter
Physics Center (IFIMAC), Universidad Aut\'onoma de
    Madrid, 28049 Madrid, Spain}


\begin{document}

\begin{abstract}
We  investigate the quantum-optical properties of the light
emitted by a nanoparticle-on-mirror cavity filled with a single
quantum emitter. Inspired by recent experiments, we model a
dark-field set-up and explore the photon statistics of the
scattered light under grazing laser illumination. Exploiting
analytical solutions to Maxwell's equations, we quantize the
nanophotonic cavity fields and describe the formation of
plasmon-exciton polaritons (or plexcitons) in the system. This
way, we reveal that the rich plasmonic spectrum of the nanocavity
offers unexplored mechanisms for nonclassical light generation
that are more efficient than the resonant interaction between the
emitter natural transition and the brightest optical mode.
Specifically, we find three different sample configurations in
which strongly antibunched light is produced. Finally, we
illustrate the power of our approach by showing that the
introduction of a second emitter in the platform can enhance
photon correlations further.
\end{abstract}

\maketitle

Surface plasmons (SPs) have been largely exploited to tailor the
classical (spatial and temporal) characteristics of the
electromagnetic (EM) fields produced by single molecules and
quantum dots~\cite{Giannini2011,Novotny2011}. Two paradigmatic
examples of such manipulation are the reshaping of their dipolar
radiation pattern by directional
nanoantennas~\cite{Taminiau2008,Curto2010} or the Purcell
reduction of their natural lifetime in
nanogaps~\cite{Farahani2005,Anger2006}. In recent years, this
ability of SPs for EM control has been also transferred into the
quantum arena~\cite{Tame2013,Marquier2017}. Initial efforts
focused on the imprinting of nonclassical features, such as
entanglement~\cite{Moreno2004}, quadrature
squeezing~\cite{Huck2009}, or sub-Poissonian
statistics~\cite{DiMartino2012}, in plasmonic waves through the
incident, driving fields. In this context, quantum emitters (QEs)
were used as the optical sources that allowed the near-field
launching of confined single plasmons in metallic
nanowires~\cite{Chang2006,Akimov2007}.

The quest for plasmon-assisted generation of radiative quantum
states of light, propagating in free-space and into the far-field,
has attracted much attention lately~\cite{FernandezDominguez2018}.
Devices based on guiding geometries decorated with in- and
out-coupling elements have been thoroughly investigated
~\cite{Bermudez2015,Kumar2021}. SPs suffer heavily from
metallic absorption in these extended systems. For this reason,
nanocavities have emerged as an alternative for nonclassical light
sources of smaller dimensions. Importantly, these nanostructures
also make it possible to fully harness the large density of
photonic states associated with SPs~\cite{Baumberg2019}.
Theoretical studies have shown that the weak interaction between a
single QE and a metallic nanosphere gives rise to moderate photon
antibunching and reduction of quantum-optical
fluctuations~\cite{Waks2010,Ridolfo2010,MartinCano2014,Liberal2018}.
Accordingly, the measurement of second-order correlation functions
below unity is taken as proof of the single emitter operation in
experiments on the Purcell effect in plasmonic
antennas~\cite{Hoang2016,Singh2018}.

The realization of stronger quantum nonlinearities with larger
near-to-far-field transfer efficiencies requires nanocavity-QE
samples that function in the strong-coupling
regime~\cite{Torma2014}. This leads to the formation of hybrid
SP-QE states, usually termed as
plexcitons~\cite{Manjavacas2011,Esteban2014,Li2016}, whose
properties can be tuned through the admixture of the interactive
character of excitons and the coherence and ubiquity of photons.
This phenomenon does not only offer new avenues for light
generation, but also lies at the core of the emergent field of
polaritonic chemistry~\cite{Feist2017,GarciaVidal2021}. The
accurate description of the signature of SP-QE interactions in
far-field optical signals relies on the EM quantization in
nanometer-sized, lossy structures, which is an area of intense
activity at the
moment~\cite{Lalanne2018,Franke2019,Tserkezis2020,Medina2021}.
Concurrently, plexciton formation in various QE-nanocavity
platforms have been realized
experimentally~\cite{Chikkaraddy2016,Santhosh2016,Gross2018,Leng2018,Ojambati2019},
and a number of theoretical models have investigated the emergence
of photon correlations in these
systems~\cite{SaezBlazquez2017,Peyskens2018,Rousseaux2020,You2020}.

In this Letter, we investigate theoretically the quantum optical
properties that plexciton strong coupling induces in the light
scattered by a plasmonic
cavity~\cite{tserkezis2015,Li2018} in a dark-field set-up~\cite{Knight2010,Lei2012}. Through radiative-corrected
quasi-static EM
calculations~\cite{CuarteroGonzalez2018,CuarteroGonzalez2020}, we
describe the near-field and radiative characteristics of the SP
modes sustained by the structure, as well as their interaction
with a molecule placed at its gap. We characterize first the
response of the bare cavity under grazing laser excitation.
Secondly, we describe the far-field intensity and second-order
correlation spectra for the configuration most explored
experimentally~\cite{Chikkaraddy2016,Santhosh2016,Leng2018}: QE at
resonance with the brightest, dipolar (lowest in frequency) SP
mode. We perform next a comprehensive study of the dependence of
photon correlations on the detuning between QE and laser
frequencies, as well as on the cavity gap size. Thus, we reveal three
different parameter ranges in which strong antibunching can be
attained.
%After assessing the impact of molecule dephasing and
%non-radiative decay, which allows us to gain insight into the
%nature of photon correlations,
Finally, we illustrate the power of our
approach by introducing a second QE in the system. We find that
the second-order correlation function can be further reduced this
way~\cite{SaezBlazquez2018}, thanks to the emergence of new
pathways for destructive quantum interference in the plexciton
ladder.

\section{Theoretical Modelling}

\begin{figure}[t]
\centering
\includegraphics[width=0.65\linewidth]{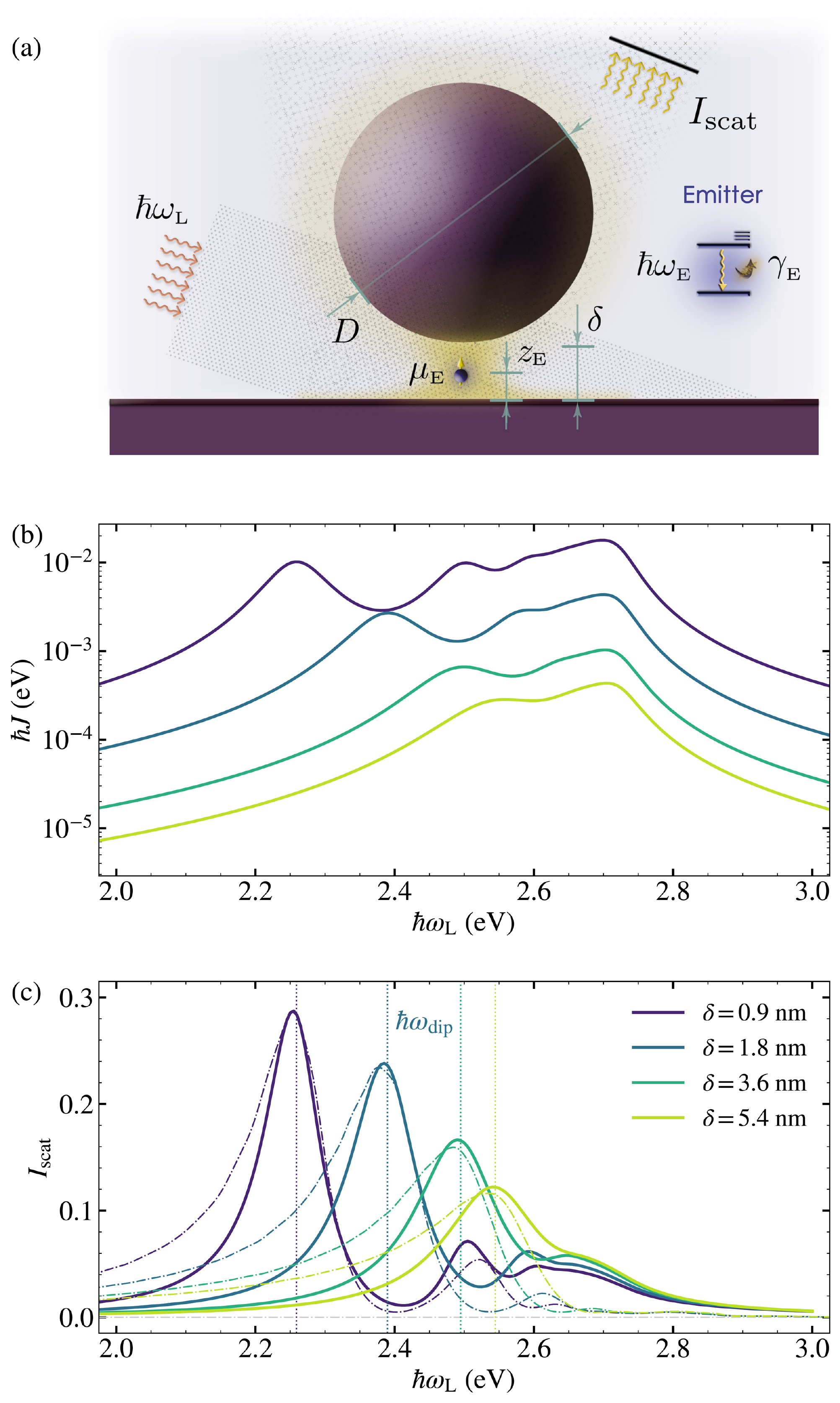}
\caption{(a) Sketch of the system, composed of a single QE coupled to the
SP fields within the gap of a NPoM cavity in a dark-field-like
set-up. The inset shows the two-level scheme modelling the emitter.
(b) Spectral density $J(\omega)$ at the gap center for different
values of the gap size $\delta$ (with $D=30$ nm). (c) Normalized
dark-field scattering spectra for the bare cavities above. Solid
lines represent $I_{\rm SP}$ given by \autoref{eq:ISP}, while
dash-dotted lines correspond to scattered intensity obtained by
numerical EM calculations. Vertical dotted lines indicate the
position of the lowest-order (dipolar) plasmon mode, with energy
$\hbar\omega_{\rm dip}$ for each gap size.}
\label{fig:modelScheme}
\end{figure}

\autoref{fig:modelScheme}(a) sketches the system of interest: an
archetypal nanoparticle-on-mirror (NPoM) cavity, formed by a $D=30$ nm diameter nanosphere
separated by a few-nanometer gap from a planar
substrate~\cite{Chikkaraddy2016,Li2018}. Both are  metallic, with
permittivity given by a Drude fitting for silver,
${\epsilon(\omega)=\epsilon_{\infty}-\omega_{\rm
p}^2/\omega(\omega+ \ii\gamma_{\rm m})}$, where $\hbar\omega_{\rm p}
= 8.91$ eV, $\epsilon_{\infty} = 9.7$ and $\hbar\gamma_{\rm m} =
0.06$ eV. For simplicity, the background refractive index is set
to unity. We employ an analytical, two-dimensional model that we
recently developed (see Ref.~\cite{CuarteroGonzalez2020} for more
details)   to describe the SP modes sustained by this geometry
(fully defined by the diameter $D$ and gap size $\delta$). This
tool is based on quasi-static solutions to Maxwell's equations
and is refined by means of the so-called radiative-reaction
correction~\cite{Carminati2006,Aubry2010}, yielding an excellent
agreement with numerical EM simulations. A QE is placed in the
NPoM gap. It is characterized by its transition dipole moment,
$\mu_{\rm E}$, transition frequency, $\omega_{\EE}$, radiative,
$\gamma_{\rm rad}$, and non-radiative, $\gamma_{\rm nrad}$, decay
rates, as well as its dephasing rate, $\gamma_{\rm deph}$. Note
that the QE radiative decay rate is simply $\gamma_{\rr} =
\omega_{\EE}^3 \mu_{\EE}^2 /(3 \pi \epsilon_0 \hbar c^3)$
\cite{bookNovotny2012} (where $\epsilon_0$ is the vacuum
permittivity and $c$ the speed of light) and its nonradiative
decay is set by its intrinsic quantum yield ${\rm QY}
=\gamma_{\rr}/(\gamma_{\rr} + \gamma_{\nr})$. The hybrid NPoM-QE
sample is driven by a grazing laser field of frequency
$\omega_{\LL}$ and amplitude $E_{\LL}$, mimicking a
dark-field-like illumination.

Our quasi-static treatment allows the labelling of the NPoM modes
in terms of two quantum numbers, their azimuthal order,
$n=1,2,3\ldots$, and the odd/even parity of their associated EM
fields across the gap center, $\sigma=\pm 1$~\cite{Aubry2010}.
Here, to simplify the notation, we combine both in a single index
$\alpha=\{n,\sigma\}$. Our theory also yields their natural
frequencies, $\omega_{\alpha}$, and broadening,
$\gamma_{\alpha}=\gamma_{\rm m}+\gamma_{\alpha}^{\rm rad}$. Note
that plasmonic absorption is the same for all SPs, given by the
Drude damping, and their radiative decay is proportional to the
square of their dipole moment, $\gamma_{\alpha}^{\rm
rad}\propto\mu_{\alpha}^2$, but the usual free-space expression
has to be corrected by the image charge distribution induced in
the metal substrate~\cite{CuarteroGonzalez2020}. We also obtained
closed expressions for the QE-SP coupling strengths, $g_{\alpha}$,
and their dependence on the QE parameters (position and natural
frequency).

In the rotating frame~\cite{thesisGardiner1995}, set by the laser
frequency $\omega_{\LL}$, and under the rotating-wave
approximation~\cite{bookLoudon2000}, the Hamiltonian for the set-up
in \autoref{fig:modelScheme}(a) is~\cite{SaezBlazquez2018}
\begin{eqnarray}
\hat{H}&=&\hbar(\omega_{\EE}-\omega_{\LL})\hat{\sigma}^{\dag}\hat{\sigma}+\sum_{\alpha}\hbar(\omega_{\alpha}-\omega_{\LL})\hat{a}^{\dag}_{\alpha}\hat{a}_{\alpha}+\notag
\\
&&+\sum_{\alpha}\hbar
g_{\alpha}(\hat{\sigma}^{\dag}\hat{a}_{\alpha}+\hat{\sigma}\hat{a}_{\alpha}^{\dag})+\hbar\Omega_{\EE}(\hat{\sigma}^{\dag}+\hat{\sigma})+\notag
\\
&&+\sum_{\alpha}\hbar\Omega_{\alpha}(\hat{a}_{\alpha}^{\dag}+\hat{a}_{\alpha}),\label{eq:Hamiltonian}
\end{eqnarray}
where $\hat{a}_{\alpha}$ and $\hat{\sigma}$ are the SP and QE
annihilation operators, respectively. Note that the third term in
\autoref{eq:Hamiltonian} describes the light-matter coupling,
where $\hbar g_\alpha = \mathbf{E}_\alpha^{(1)} \cdot
\boldsymbol{\mu}_{\EE}$ and $\mathbf{E}_\alpha^{(1)}$ is the
quantized one-photon field strength of mode $\alpha$ at the QE
position. Then,
$\hbar\Omega_{\EE}=\boldsymbol{E}_{\LL}\cdot\boldsymbol{\mu}_{\EE}$
and
$\hbar\Omega_{\alpha}=\boldsymbol{E}_{\LL}\cdot\boldsymbol{\mu}_{\alpha}$
are the coherent pumping amplitudes. Note that dipole moments and
incident fields are oriented vertically in
\autoref{fig:modelScheme}(a). The master equation for the
steady-state of the system including SP damping and QE decay and
dephasing reads
\begin{eqnarray}
&&\frac{\ii}{\hbar}[\hat{\rho},\hat{H}]+\sum_{\alpha}\frac{\gamma_{\alpha}}{2}\mathcal{L}_{\hat{a}_{\alpha}}[\hat{\rho}]+\notag \\
&&+\frac{\gamma_{\rm rad}+\gamma_{\rm
nrad}}{2}\mathcal{L}_{\hat{\sigma}}[\hat{\rho}]+\frac{\gamma_{\rm
deph}}{2}\mathcal{L}_{\hat{\sigma}^\dag\hat{\sigma}}[\hat{\rho}]=0,
\label{eq:mastereq}
\end{eqnarray}
where the Lindblad terms have the usual form
$\mathcal{L}_{\hat{\mathcal{O}}} = 2 \hat {\mathcal{O}} \hat \rho
\hat {\mathcal{O}}^\dagger - \hat {\mathcal{O}}^\dagger \hat
{\mathcal{O}} \hat \rho - \hat \rho \hat {\mathcal{O}}^\dagger
\hat{\mathcal{O}}$.

\section{Results and discussion} \label{sec:Results}

Before investigating far-field optical signatures of light-matter
interactions in the hybrid QE-SP system, we employ our theory to
characterize the bare plasmonic cavity first. We compute the
spectral density weighting the local density of photonic states
and the QE-SP coupling strength at the NPoM gap. We have recently
shown that this is given by~\cite{Medina2021}
\begin{eqnarray}
J(\omega)&=&\frac{\hbar}{\pi}{\rm Im} \left\{
\sum_{\alpha}g_{\alpha}\left(\tilde{\mathbf{H}}-\hbar\omega\right)^{-1}_{\alpha\alpha}
g_{\alpha} \right\}\notag \\
&=&\sum_{\alpha} \frac{g^2_{\alpha}}{\pi}
\frac{\gamma_{\alpha}/2}{(\omega - \omega_{n,\sigma})^2 +
\gamma^2_{\alpha}/4},
\end{eqnarray}
where
$\tilde{\mathbf{H}}_{\alpha\beta}=\hbar\left(\omega_{\alpha}-\ii
\tfrac{\gamma_{\alpha}}{2}\right)\delta_{\alpha\beta}$ is equal to
the coefficient matrix of the SP modes in the effective
non-Hermitian Hamiltonian governing the coherent evolution in the
Lindblad master equation.

\autoref{fig:modelScheme}~(b) renders the spectral density at the
center of the NPoM gap, $z_{\EE} = 0.5\,\delta$ ($z=0$ corresponds
to the substrate surface), for cavities with $D=30\,$nm and
$\delta$ ranging from 0.9 (purple) to $5.4\,$nm (light green). The
coupling constants are proportional to the QE dipole moment,
$\g_\alpha\propto\mu_{\EE}$. They were evaluated at $\mu_{\EE} =
0.55$ e$\cdot$nm, the value which we will consider in our
plexcitonic systems. As previously reported~\cite{Li2016}, the
smaller the gap, the larger $J(\omega)$. In all cases, the spectra
present a low-frequency maximum originating from the brightest,
dipolar SP mode, $\alpha=\{1,1\}$, which redshifts with decreasing
$\delta$; and another maximum in the vicinity of the asymptotic
surface plasmon frequency (${\omega_{\rm
P}/\sqrt{1+\epsilon_\infty}}$) due to the pseudomode that results
from the spectral overlapping of high order SPs~\cite{delga2014}.
For small enough gap sizes, the contributions from quadrupolar and
higher order even modes (specifically, $\alpha=\{2-4,1\}$) are
also apparent.

We focus next on the far-field response of the bare NPoM
structure. We compute the scattering spectrum by solving
\autoref{eq:mastereq} removing all the QE-related terms. Once the
SP steady-state density matrix, $\hat \rho_{\rm SP}$, is known,
the far-field scattering intensity can be computed as
\begin{equation}
I_{\rm SP}= \langle \hat{E}_{\rm SP}^- \hat{E}_{\rm SP}^+
\rangle=\sum_{\alpha,\beta}\mu_\alpha\mu_\beta{\rm
tr}\{\hat{a}_\alpha^\dagger \hat{a}_\beta \hat{\rho}_{\rm
SP}\},\label{eq:ISP}
\end{equation}
where $\hat{E}_{\rm SP}^- = \sum_\alpha \mu_\alpha
\hat{a}_\alpha^\dagger$ is the (negative frequency part of the)
electric far-field operator. For simplicity, we are dropping the
Dyadic Green's function in the definition of the electric field
operator, which would account for the spatial pattern of the
cavity fields. Importantly, the cross terms in
\autoref{eq:Hamiltonian} reflect the emergence of superposition
effects in the photon emission from different SP modes.

\autoref{fig:modelScheme}(c) plots the scattering spectra for the
NPoM configurations in \autoref{fig:modelScheme}(b). Dash-dotted
lines correspond to COMSOL Multiphysics$^{\rm TM}$ simulations of
the scattering efficiency (defined as the cross section normalized
to physical size) in the quasi-static limit, while solid lines
plot the prediction from \autoref{eq:ISP}. Note that the latter
have been scaled vertically (by the same factor for all $\delta$)
to facilitate the comparison between both sets of data. We can
observe that the spectra are governed by a large peak at laser
frequencies in the vicinity of the lowest, dipolar SP
($\alpha=\{1,1\}$). The condition $\omega_{\LL}=\omega_{\rm dip}$
is indicated by vertical dotted lines in all cases. Higher order
SP maxima are also evident, specially at small $\delta$. On the
contrary, there is not any pseudomode signature in the scattering
signal, as expected from the dark character of the SP modes that
form it. The NPoM spectra in \autoref{fig:modelScheme}(c) present
scattering minima at laser frequencies between SP resonances.
These are the so-called invisibility dips, which emerge (more
clearly in log scale) due to the destructive interference in the
photon emission by different plasmonic channels~\cite{Aubry2010}.

\begin{figure}[tb]
\centering
\includegraphics[width=\linewidth]{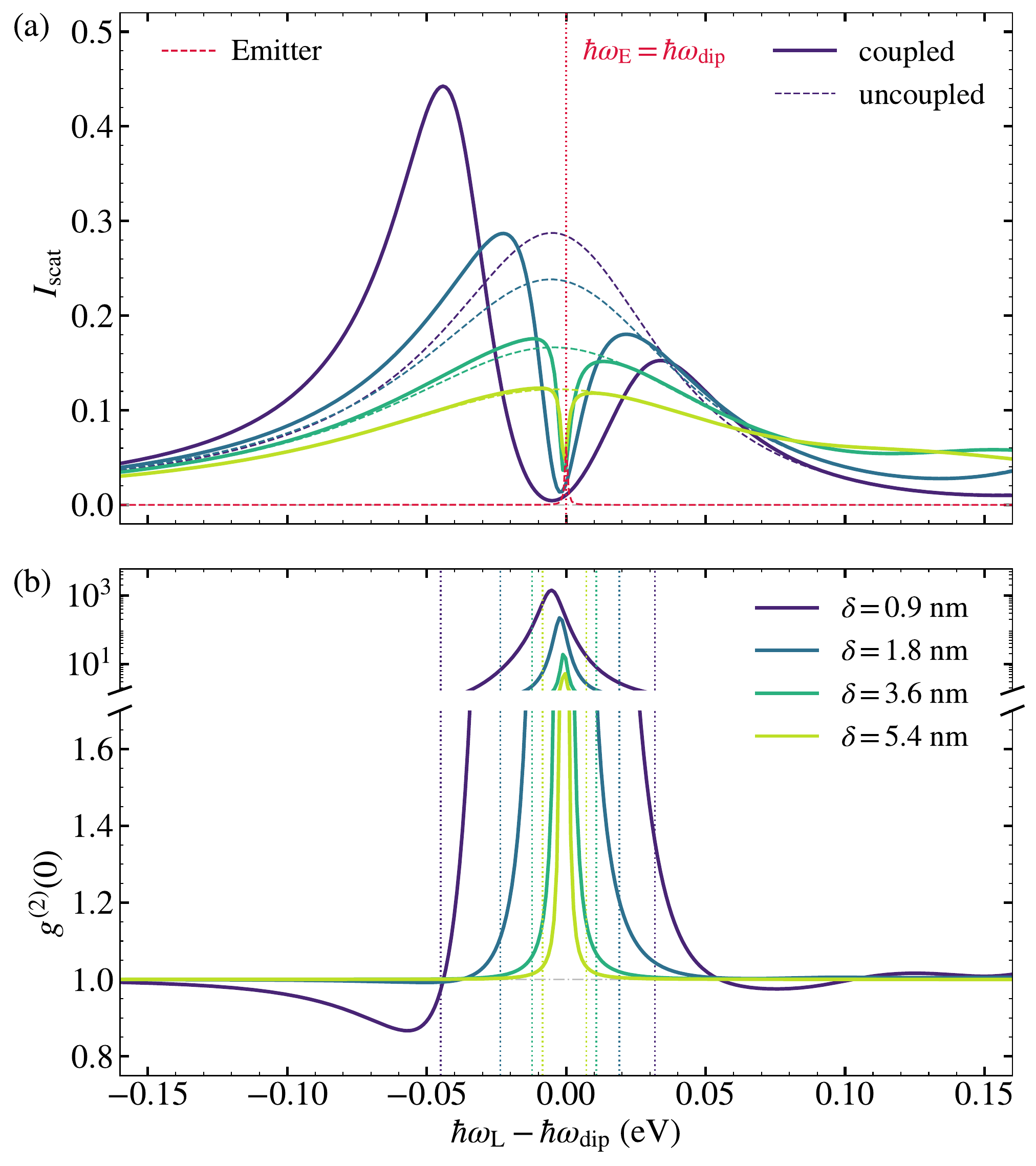}
\caption{(a) Far-field spectra for the bare cavities in
\autoref{fig:modelScheme} (dashed lines) and the hybrid SP-QE
samples (solid lines) that result from introducing a
vertically-oriented molecule at the gap center. $I_{\rm scat}$ is
plotted against the detuning between the incident laser and the
dipolar SP, $\hbar \omega_{\LL} - \hbar \omega_{\rm dip}$, and
only in the vicinity of this cavity mode. The dashed red line
represents the free-standing QE spectrum, which is, in all
cases, at resonance with the dipolar SP, $\omega_{\EE}=\omega_{\rm
dip}$ (vertical dotted line). (b) Zero-delay second-order
correlation function, $g^{(2)}(0)$, versus laser detuning for the
same NPoM-QE configurations in (a). Vertical dotted lines indicate
the plexciton frequencies in the one-excitation manifold for each
gap size.}\label{fig:dipolar}
\end{figure}

Next, we place a vertically-oriented QE at the center of the gap
of the NPoM geometries in \autoref{fig:modelScheme}. Reproducing
previous experimental setups, we set the QE frequency at resonance
with the dipolar SP, $\omega_{\EE} = \omega_{\rm dip}$, which is
different for each $\delta$. This way, the signature of QE-SP
interaction is expected to be most apparent in the far-field. We
take QY=0.65, in agreement with values reported for molecular
dyes, such as Atto 647N~\cite{Ojambati2019}. The associated
radiative and nonradiative decay rates are therefore in the
$10^{-6}-10^{-7}$ eV range (note that these depend on
$\omega_{\EE}$). Additionally, we consider a QE dephasing rate of
$\gamma_{\rm deph} = 1$ meV \cite{Clear2020}. The spectra for the
hybrid NPoM-QE system can be obtained from the steady-state
density matrix solution, $\hat{\rho}$, for the full master
equation in \autoref{eq:mastereq},
\begin{equation}
I_{\rm scat}= \langle \hat{E}_{\rm scat}^- \hat{E}_{\rm scat}^+
\rangle={\rm tr}\{\hat{E}_{\rm scat}^- \hat{E}_{\rm
scat}^+\hat{\rho}\},\label{eq:Iscatt}
\end{equation}
where in order to account for the open character of the plasmonic
cavity, the electric field operator, ${\hat{E}_{\rm scat}^-
=\hat{E}_{\rm SP}^-+\mu_{\EE}\hat{\sigma}^\dagger}$, includes now the
emission from the molecule itself~\cite{SaezBlazquez2018}. Note
that we can also compute the density matrix, $\hat \rho_{\rm E}$,
and scattering intensity for the free-standing emitter, obtained from
the QE terms in \autoref{eq:mastereq} and $I_{\rm E}= \mu_{\rm
E}^2{\rm tr}\{\hat{\sigma}^\dagger \hat{\sigma} \hat\rho_{\rm
E}\}$.

\autoref{fig:dipolar}(a) shows the scattering intensity versus
$\hbar(\omega_{\rm L}-\omega_{\rm dip})$, the laser detuning with
respect to the dipolar SP, which allows the direct comparison
between different cavities. Dashed lines correspond to the
Lorentzian-like spectral profile of $I_{\rm SP}$ for all
structures, while solid lines plot the spectra for the plexcitonic
samples. For reference, the red dotted line shows $I_{\rm E}$
normalized to the nanoparticle size, whose linewidth is given by
$\gamma_{\rm E}=\gamma_{\rr}+\gamma_{\nr}+\gamma_{\rm deph}\simeq
1$ meV. For large gaps, and therefore lower QE-SP coupling
strengths, the presence of the molecule leads to the appearance of
a scattering dip at $\omega_{\rm L}=\omega_{\rm E}$ of width
similar to $\gamma_{\rm E}$. This phenomenology, closely related to
the electromagnetic induced transparency, is in accordance
with that reported previously for single metallic nanoparticles in
the weak-interaction regime~\cite{Ridolfo2010}. For small
$\delta$, the far-field spectra develop a well-defined Rabi
doublet lineshape. This is the fingerprint of the onset of strong
coupling between the bright plasmon mode and the
molecule~\cite{Chikkaraddy2016}. These two scattering maxima are
originated from the upper and lower plexcitonic states that have
been formed in the cavity (see below). Two different mechanisms
contribute to make the intensity of the lower plexciton larger
than the upper one. On the one hand, the former (latter) results
from the constructive (destructive) interference of the SP and QE
emission channels~\cite{SaezBlazquez2018}. On the other hand, it
has been shown that, despite being highly detuned, higher
frequency, neighboring SP modes can also increase the Rabi
asymmetry in these systems~\cite{CuarteroGonzalez2020}.

Our approach enables us to characterize the light scattered by the
NPoM-QE system beyond the intensity spectra above. We can employ
it to analyze the scattered photon statistics through the
so-called zero-delay second-order correlation
function~\cite{bookLoudon2000}
\begin{equation}
g^{(2)}(0) = \langle \hat {E}_{\rm scat}^- \hat {E}_{\rm scat}^-
\hat{E}_{\rm scat}^+ \hat{E}_{\rm scat}^+ \rangle / I_{\rm scat}^2
\ ,
\end{equation}
which gives the probability of detecting two coincident photons in
the far-field. Although not discussed above, the bosonic character
of SPs yields $g^{(2)}(0) = 1$ for the bare NPoM cavities in
\autoref{fig:modelScheme}. In the following, we will explore the
conditions in which the plexcitonic system deviates from these
Poissonian statistics, with special focus on the emergence of
negative correlations, antibunching or sub-Poissonian statistics.
With all these terms we will refer to photon emission
characterized by a second-order correlation function below unity,
$g^{(2)}(0) < 1$.

In \autoref{fig:dipolar}(b), we plot the zero-delay second-order
correlation function for the NPoM-QE samples in panel (a).
Vertical dotted lines indicate the two plexciton frequencies in
the first manifold for all geometries (which coincide with the
scattering intensity maxima~\cite{SaezBlazquez2017}). We can
observe that $g^{(2)}(0)\gg 1$, bunched emission or more
rigorously, super-Poissonian statistics, takes place between them.
The maximum in $g^{(2)}(0)$ occurs at $\omega_{\rm
L}\simeq\omega_{\rm E}$ and redshifts and increases with
decreasing gap size (larger QE-SP coupling). Only for $\delta=0.9$
nm (purple line) negative photon correlations are apparent. A
region of moderate antibunching, $g^{(2)}(0)> 0.8$, develops for
laser frequencies slightly below the lower plexciton frequency
(note that an even shallower dip also occurs at $\omega_{\rm L}$
above the upper plexciton). The correlation spectra overlap with
those obtained by neglecting photon emission by SP modes different
from the dipolar one, which therefore do not play any role in this
particular NPoM-QE configuration.

\begin{figure*}[htb]
\centering
\includegraphics[width=\linewidth]{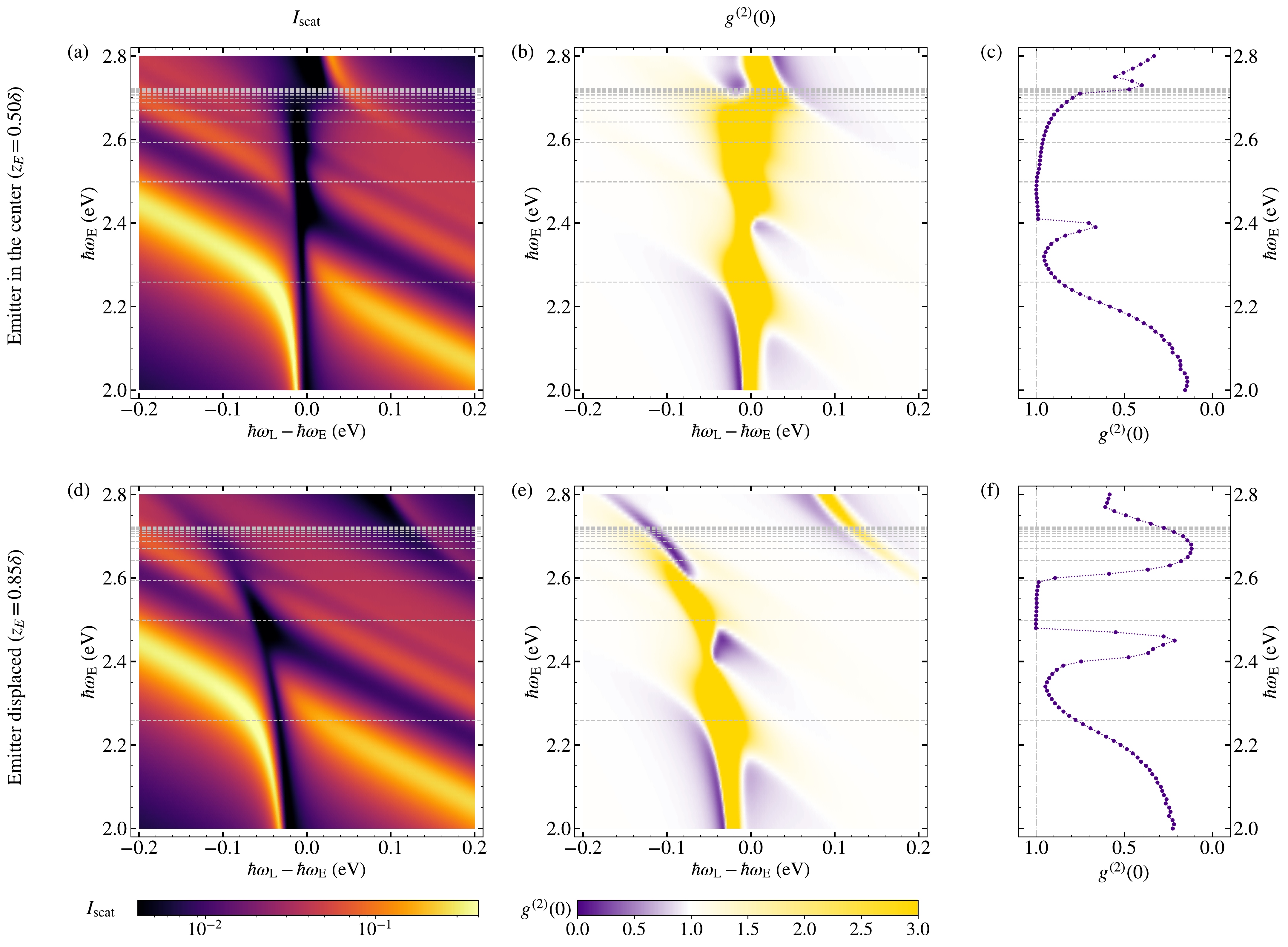}
\caption{Scattering intensity $I_{\rm scat}$ (first column) and
second-order correlation function $g^{(2)}(0)$ (second column)
versus laser-QE detuning $\hbar \omega_{\LL} - \hbar \omega_{\EE}$
and QE frequency $\hbar\omega_{\EE}$ for emitter at the center of
the gap (top, a-c) or vertically displaced (bottom, d-f). Panels in third column plot the minimum of $g^{(2)}(0)$ as a function
of the emitter
frequency extracted from (b) and (e). In all panels, horizontal
dotted lines indicate the position of the NPoM SP frequencies.}
\label{fig:contour}
\end{figure*}

The negative correlations observed in \autoref{fig:dipolar}(b) can
be attributed to the so-called \textit{photon blockade
effect}~\cite{Imamoglu1997,Birnbaum2005}, where the presence of an
excitation in the system prevents the absorption of a second
photon of the same frequency due to the anharmonicity of the
plexciton ladder. This phenomenon becomes stronger as the
light-matter interaction strengthens, which means that smaller gap
sizes or larger QE dipole moments would be required to reduce
$g^{(2)}(0)$ further. However, there exists another effect yielding
sub-Poissonian photon emission, known as interference-induced or
\textit{unconventional
antibunching}~\cite{Carmichael1991,Rempe1991}. Thoroughly analyzed
in single-mode semiconductor
microcavities~\cite{Radulaski2017}, it develops
only when the driving laser is far from resonance, and due to
destructive quantum interference among different de-excitation
pathways in the system. In the following, we investigate the
emergence of both antibunching mechanisms in our plexcitonic
samples, exploring the full richness of the NPoM spectrum through
the emitter and laser frequencies and the emitter position.

\autoref{fig:contour} shows intensity (left panels) and
second-order correlation function (central panels) maps as a
function of the detuning of the laser with respect to the QE
frequency (horizontal axes) and the emitter frequency itself
(vertical axes). The gap size is fixed to $\delta=0.9$ nm, and two
different emitter positions are considered: at the center of the
gap, $z_{\rm E}=0.5\delta$ (top, a-c), and displaced vertically
towards the nanoparticle surface, $z_{\rm E}=0.85\delta$ (bottom,
d-f). In these panels, the SP frequencies, $\hbar\omega_\alpha$, are
marked by horizontal dotted lines. Note that the purple solid
lines in \autoref{fig:dipolar} correspond to horizontal cuts of
\autoref{fig:contour}(a) and (b) in the vicinity of the dipolar
SP. In this range of QE frequencies and below it ($\omega_{\EE}
\lesssim \omega_{\rm dip}=2.26$ eV), $I_{\rm scat}$ develops a
clear Rabi doublet lineshape, associated with the two plexcitons
that emerge from the strong coupling of QE and dipolar SP. For
red-detuned QEs, the lower (upper) plexciton has a more
emitter-like (plasmon-like) character, and its position approaches
$\omega_{\EE}$ ($\omega_{\rm dip}$). On the contrary, for blue
detuned QEs, the signature of higher order SPs becomes apparent,
and $I_{\rm scat}$ reproduces a similar anticrossing phenomenology
as that around the dipolar SP mode. The intensity maps for both
emitter positions are similar, with a remarkable difference: while
the scattering dip between upper and lower plexcitons is always at
$\omega_{\rm L}=\omega_{\EE}$ at the gap center (a), it redshifts
with increasing QE frequency for $z_{\EE}=0.85\delta$ (d). This is a
direct consequence of the large coupling to the plasmonic
pseudomode that the emitter experiences when it is placed in close
proximity to the nanoparticle boundary. This is evident in the
far-field spectra even for $\omega_{\EE}$ significantly detuned
from the pseudomode frequency ~\cite{CuarteroGonzalez2020}.

The photon correlation maps in \autoref{fig:contour}(b) and (d)
show that $g^{(2)}(0)$ has a higher sensitivity on the QE position
than $I_{\rm scat}$. Both panels expose that bunching emission
(yellow, $g^{(2)}(0)>1$) takes place at the conditions for
plexciton anticrossing, where $I_{\rm scat}$ is minimum. They also
reveal that much stronger negative correlations than the resonant
($\omega_{\EE}=\omega_{\rm dip}$) configuration considered in
\autoref{fig:dipolar} can be achieved by exploiting the full
plasmonic spectrum of NPoM cavities. To clarify the degree of
antibunching attainable in these systems, \autoref{fig:contour}(c)
and (f) plot the spectral minimum of $g^{(2)}(0)$ as a function of
the QE frequency. For both $z_{\EE}$, a region of sub-Poissonian
statistics is apparent at emitter frequencies below the dipolar
SP, which becomes stronger and spectrally broader for lower
$\omega_{\EE}$. As we discussed above, at laser frequencies
slightly below the emitter frequency, these negative correlations
are generated via the photon blockade effect, yielding
$g^{(2)}(0)$ values below 0.2. On the contrary, a weaker
interference-induced antibunching takes place in this region but
for $\omega_{\rm L}>\omega_{\rm E}$, see panels (b) and (e).

Another region yielding $g^{(2)}(0)<1$ in
\autoref{fig:contour}(b) and (e) occurs at QE frequencies
approaching the pseudomode. Negative correlations are weaker than
below the dipolar SP and, as discussed below, they have a
different, interference-induced, origin. Note that these become
more apparent for $z_{\rm E}=0.85\delta$, since the coupling to
the higher-energy plasmon modes increases this way \cite{Li2016}.
Antibunching also takes place in a third NPoM-QE configuration, at
QE frequencies in between the dipolar (lowest) and the quadrupolar
(second lowest, $\alpha=\{2,1\}$) SP cavity modes (2.3 eV
$\lesssim\hbar\omega_{\rm E}\lesssim 2.5$ eV)---exactly at the
parameter range where a scattering (invisibility) dip, due to
destructive interference effects in the emission by these two SPs,
evolves in $I_{\rm scat}$~\cite{Aubry2010}. We can therefore
conclude that this phenomenon does not only emerge in the
intensity spectrum, but also in the photon statistics. The QE
position weights the relative coupling between the emitter and
both cavity modes, and thus the strength of the interference
that suppresses two-photon processes, which seems to be larger
(reaching $g^{(2)}(0)$ below 0.2) for displaced emitters.

\begin{figure}[t]
\centering
\includegraphics[width=\linewidth]{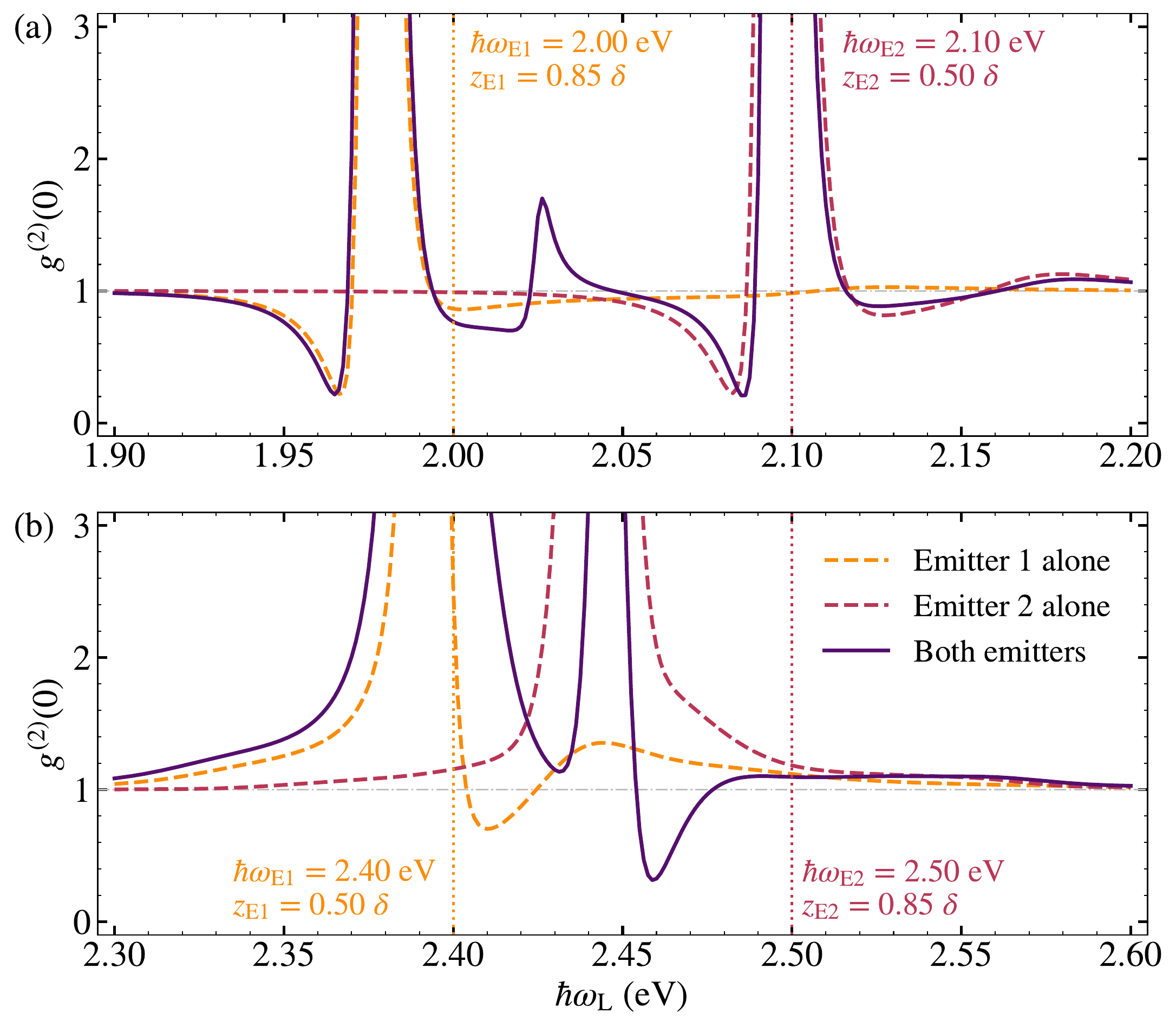}
\caption{Second-order correlation function versus laser frequency
$\omega_{\LL}$ for two different plexcitonic systems, containing
two QEs (one at the gap center, another vertically displaced). The
QE frequencies are either red-detuned with respect to the dipolar
plasmon (a) or lying within the scattering (invisibility) dip (b).
Here, $g^{(2)}(0)$ for the two-emitter configurations (continuous
lines) are compared against the corresponding single-emitter
calculations (dashed lines). Vertical dotted lines indicate the
values of $\hbar\omega_{\EE}$ in each case.}\label{fig:zwei}
\end{figure}

Lastly, we investigate whether, as previously reported for
single-mode cavity models~\cite{SaezBlazquez2017}, the presence of
a second emitter may be beneficial for the generation of
nonclassical light in QE-SP systems. We consider two
vertically-oriented emitters hosted in the small gap cavity above
($\delta=0.9$ nm, $D=30$ nm).
%The nonradiative and dephasing rates
%are set back to their original values, and
The two QE positions
are chosen to be the same as in \autoref{fig:contour}: $z_{\EE} =
0.50 \delta$ and $z_{\EE} = 0.85 \delta$. \autoref{fig:zwei} plots
the second-order correlation function versus the laser frequency
for two different configurations, chosen from the single-emitter
samples that yield sub-Poissonian emission in the same figure. In
panel (a), both emitters are red-detuned with respect to the
dipolar SP, while they are at the invisibility dip between the
dipolar and quadrupolar modes in panel (b). Continuous lines plot
$g^{(2)} (0)$ for the two emitters, while dashed lines represent
the associated single-emitter cases. In both panels, we consider
QE frequencies slightly separated, $\hbar(\omega_{\rm
E2}-\omega_{\rm E1})=0.1$ eV, which is of the order of the Drude
damping, $\gamma_{\rm m}$. The correlation function for further
QE-QE detunings is basically the superposition of the two
single-emitter calculations. On the other hand, if $\omega_{\EE}$
is the same for both emitters, the plexciton emission is that of
the single QE with a larger transition dipole moment.
\autoref{fig:zwei} explores the intermediate regime: a significant
enhancement of negative correlations is not apparent at low QE
frequencies (a), but a strong reduction in $g^{(2)} (0)$ takes
place for $\omega_{\rm E1}$ and $\omega_{\rm E2}$ at the
invisibility dip. In particular, we observe a dip in the
correlation function at laser frequencies between the two QE
lines. In that minimum, $g^{(2)} (0) \sim 0.3$, while the
corresponding single-emitter spectra do not present values below
0.7. Thus, we can conclude that indeed the interplay and
interaction between various SP modes and QEs can be exploited to
enhance the degree of antibunching in the photon emission by NPoM
plexcitonic systems.

\section{Conclusion}

We have presented a master equation description of the far-field
photon emission by a plasmonic nanoparticle-on-mirror cavity
strongly coupled to a single molecule or quantum emitter.
Parameterized through classical electromagnetic calculations, we
have employed our model to characterize the classical and quantum
optical properties of the light scattered by this hybrid system in
a dark-field-like set-up. First, we have found that the formation
of plexcitons does not yield significant antibunching in the most
explored sample configuration, in which the molecular transition
is at resonance with the dipolar cavity mode. Next, by varying the
laser and emitter frequencies, we have explored the whole
plasmonic spectrum of the nanostructure. This way, we have found
that large negative photon correlations take place at three
different emitter frequencies: below the dipolar plasmon, at the
invisibility dip between this mode and the quadrupolar one, and at
resonance with the plasmonic pseudomode.
%Through the modification of the molecule nonradiative decay and
%dephasing rates, we have identified the physical mechanism behind
% the sub-Poissonian statistics in each case.
Finally, we have demonstrated that, under certain conditions,
photon antibunching can be enhanced through the introduction of a
second molecule in the nanocavity. We believe that our theoretical
findings shed light into recent experiments, and can serve as a
guide for the design of devices for quantum light generation
through the strong coupling of light and material states at the
nanoscale.

\section{Acknowledgment}

This work has been sponsored by the Spanish
MCIN/AEI/10.13039/50110001033 and by "ERDF A way of making Europe"
through Grant Nos. RTI2018-099737-B-I00 and CEX2018-000805-M
(through the Mar\'ia de Maeztu program for Units of Excellence in
R$\&$D). We also acknowledge funding from the 2020 CAM Synergy
Project Y2020/TCS-6545 (NanoQuCo-CM). R.S.B. acknowledges
support by the Austrian Science Fund (FWF) through
Grant No. P31701 (ULMAC).

\begin{tocentry}
\begin{center}
\includegraphics[angle=0,width=0.65\textwidth]{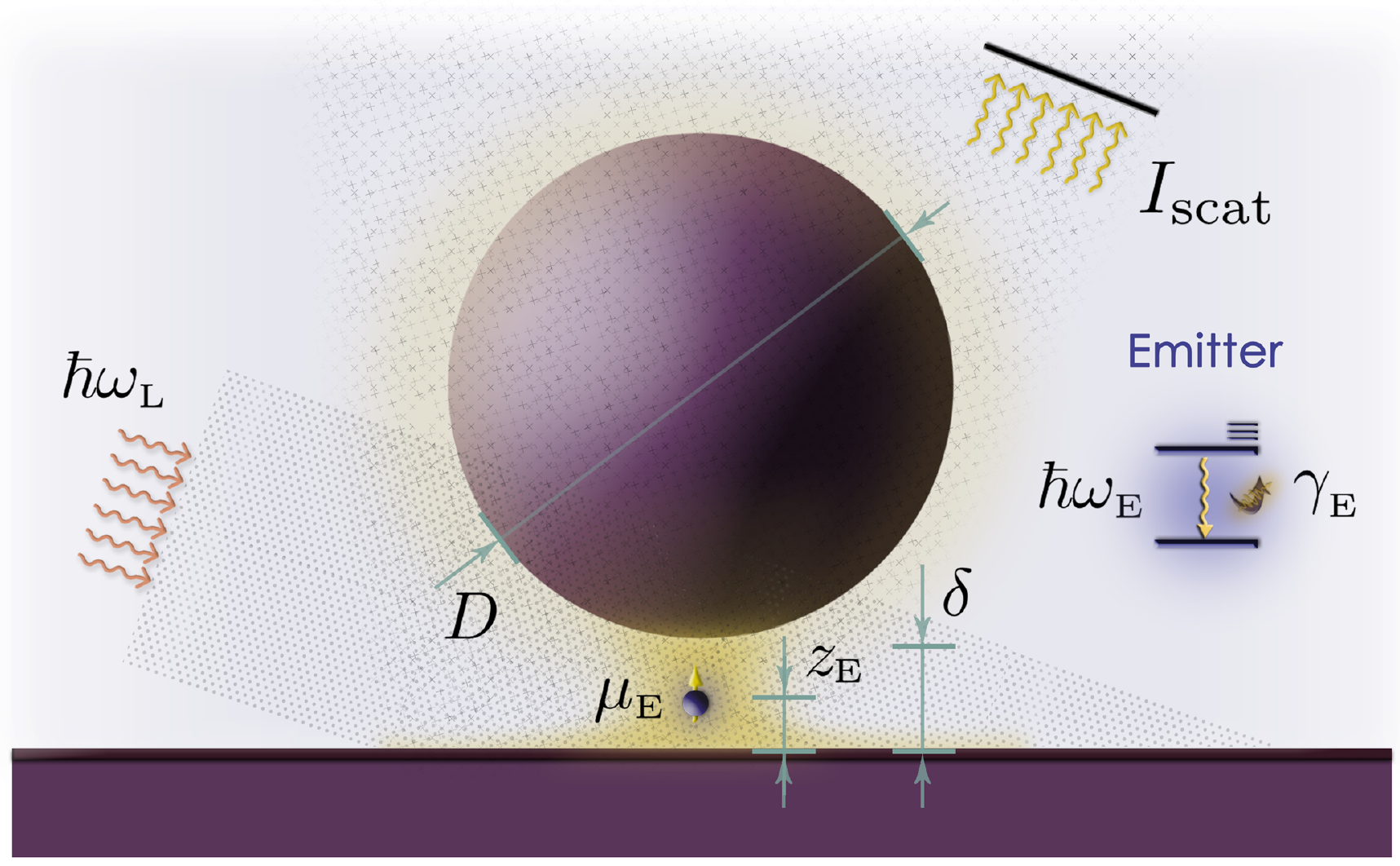}
\end{center}
\end{tocentry}

\bibliography{bibCaval} % Tell bibtex which .bib file to use (this one is some example file in TexLive's file tree)
\bibliographystyle{achemso} % Tell bibtex which bibliography style to use

\end{document}